\documentclass[11pt,a4paper]{article}
\pdfoutput=1
\usepackage{jheppub}
\usepackage{etoolbox}
\usepackage{graphicx}
\usepackage{subcaption}
\usepackage[dvipsnames]{xcolor}
\usepackage{dsfont}
\makeatletter
\patchcmd{\maketitle}{\@fpheader}{\\}{}{}
\makeatother

\title{Traversable Asymptotically Flat Wormholes with Short Transit Times}

\author[]{Zicao Fu,}
\author[]{Brianna Grado-White,}
\author[]{and Donald Marolf}

\affiliation[]{Department of Physics, University of California, Santa Barbara, CA 93106, USA}

\emailAdd{zicaofu@physics.ucsb.edu}
\emailAdd{brianna@physics.ucsb.edu}
\emailAdd{marolf@physics.ucsb.edu}

\abstract{We construct traversable wormholes by starting with simple four-dimensional classical solutions respecting the null energy condition and containing a pair of oppositely charged black holes connected by a non-traversable wormhole.  We then consider the perturbative back-reaction of bulk quantum fields in Hartle-Hawking states. Our geometries have zero cosmological constant and are asymptotically flat except for a cosmic string stretching to infinity that is used to hold the black holes apart. Another cosmic string wraps the non-contractible cycle through the wormhole, and its quantum fluctuations provide the negative energy needed for traversability. Our setting is closely related to the non-perturbative construction of Maldacena, Milekhin, and Popov (MMP), but the analysis is complementary. In particular, we consider cases where back-reaction slows, but fails to halt, the collapse of the wormhole interior, so that the wormhole is traversable only at sufficiently early times. For non-extremal backgrounds, we find the integrated null energy along the horizon of the classical background to be exponentially small, and thus traversability to be exponentially fragile. Nevertheless, if there are no larger perturbations, and for appropriately timed signals, a wormhole with mouths separated by a distance $d$ becomes traversable with a minimum transit time $t_{\text{min  transit}} = d + \text{logs}$.  Thus $\frac{t_{\text{min  transit}}}{d}$ is smaller than for the eternally traversable MMP wormholes by more than a factor of 2, and approaches the value that, at least in higher dimensions, would be the theoretical minimum. For contrast we also briefly consider a `cosmological wormhole' solution where the back-reaction has the opposite sign, so that negative energy from quantum fields makes the wormhole harder to traverse.}

\begin{document}
\maketitle
\section{Introduction}
The study of wormholes in general relativity dates back many years (see e.g. \cite{Einstein:1935tc,Graves:1960zz,Morris:1988tu}), with varying discussions of whether an observer might be able to pass through and perhaps find a shortcut to a distant region.  In particular, it is now well understood that the existence of traversable wormholes is limited in two important ways. First, topological censorship theorems \cite{Friedman:1993ty,Galloway:1999bp} forbid wormholes from being traversable in globally hyperbolic solutions to Einstein-Hilbert gravity coupled to matter satisfying the null energy condition (NEC)\footnote{Though traversable wormholes can be constructed if one drops the requirement of global hyperbolicty, e.g. by introducing NUT charge \cite{Ayon-Beato:2015eca,PhysRevD.96.025021}.}, $T_{ab} k^a k^b \ge 0$. Second, even when the NEC is violated by quantum effects, general arguments expected to hold in quantum gravity forbid wormholes in globally hyperbolic spacetimes from providing the fastest causal curves between distant points \cite{Wall:2013uza,Gao:2016bin}. This condition also prohibits the further possible pathologies discussed in \cite{Morris:1988tu,Visser:1995cc,Lobo:2007zb}. Recently, several examples of traversable wormholes supported by well-controlled quantum effects and respecting the above restrictions have been constructed \cite{Gao:2016bin, Maldacena:2017axo,Maldacena:2018lmt, Maldacena:2018gjk,  Fu:2018oaq, Caceres:2018ehr}. Instantons producing such wormholes by quantum tunneling were also discussed in \cite{Horowitz:2019hgb}.
While the second limitation significantly restricts the utility of any shortcut they might provide,
such solutions remain of theoretical interest.

These various traversable wormholes solutions naturally fall into two classes. Wormholes in the first class (see e.g. \cite{Gao:2016bin, Maldacena:2017axo,Maldacena:2018lmt,Caceres:2018ehr}) connect two separate asymptotic anti-de Sitter (AdS) regions and are supported by negative energy in the bulk that is generated by explicit couplings between the two dual boundary CFTs. While such couplings are non-local and acausal from the perspective of the bulk, they may be thought of as simple models for couplings that would be induced between wormhole mouths lying in the same asymptotic region and interacting causally through ambient space. Wormholes in this second, more natural class were constructed in \cite{Maldacena:2018gjk, Fu:2018oaq}. In particular, Maldacena, Milekhin, and Popov (MPP) \cite{Maldacena:2018gjk} used a nearly-AdS$_2$ approximation to construct a static wormhole in asymptotically flat spacetime. This approach allowed \cite{Maldacena:2018gjk} to address many non-perturbative issues. 

In contrast, \cite{Fu:2018oaq} used a perturbative framework to give a general method of constructing traversable wormholes with both mouths in the same asymptotic region, and in particular argued that a broad class of (almost traversable) classical wormhole backgrounds would become traversable after incorporating the back-reaction from standard local quantum fields in Hartle-Hawking states.  By an almost traversable background, we mean one in which there is a null geodesic $\gamma$ traversing the wormhole that lies in both the boundary of the past of future null infinity and the boundary of the future of past null infinity.  As in e.g. \cite{Gao:2016bin}, under these circumstances a negative value for the integrated null stress tensor along $\gamma$ will often lead back-reaction moving $\gamma$ into both the past of future null infinity and the future of past null infinity; i.e., the wormhole becomes traversable.  Note, however, that in contrast to the wormholes of \cite{Maldacena:2018gjk}, this perturbative approach generally yields wormholes with strong time-dependence, so that the back-reaction slows but does not stop the collapse of the wormhole interior.  The result is that the wormhole is traversable only at sufficiently early times.  This is the price to be paid for studying a more general class of constructions.  Consistent with the results of \cite{Maldacena:2018gjk}, and as discussed in \cite{Fu:2018oaq} and also reviewed below, perturbative calculations indicate that the wormholes described here and in \cite{Fu:2018oaq} can in fact become time-independent in the limit where the background almost-traversable wormhole becomes extremal. 

\begin{figure}[h]
\centering
\includegraphics[width =0.35\textwidth]{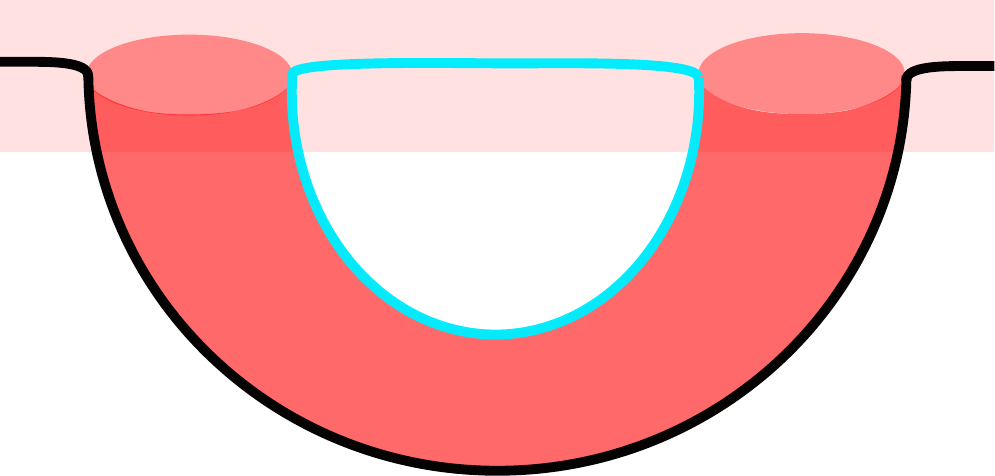}
\caption{A moment of time in a spacetime with a wormhole (shaded region) formed by adding a handle to a space with a single asymptotic region.  The wormhole is threaded by two cosmic strings, one stretching to infinity (black line) and the other compact (blue line). The string that stretches to infinity provides a tension that counteracts the gravitational (and, in our case, also electric) attraction of the two mouths, as well as the tension of the compact string, and thus prevents the black holes from coalescing. Quantum fluctuations from the compact cosmic string will render the wormhole traversable.}
\label{fig:wormhole}
\end{figure}

Here, we return to the perturbative framework of \cite{Fu:2018oaq} in order to explore the above back-reaction in more detail for a simple class of classical wormholes (suggested in \cite{Fu:2018oaq} and closely related to the setting of \cite{Maldacena:2018gjk}) which have both mouths in the same asymptotically flat region of spacetime.
Our classical backgrounds contain a pair of charged, Reissner-Nordstr\"om-like black holes held apart by the tension of a cosmic string that threads the wormhole and stretches to infinity. We also include a second cosmic string that wraps the non-contractable cycle through the wormhole; see figure \ref{fig:wormhole}. The classical wormholes are not traversable, but are almost so.  Quantum fluctuations from this compact cosmic string generate the negative Casimir energy whose back-reaction renders the wormhole traversable.  As in \cite{Fu:2018oaq}, the back-reacted wormhole will generally exhibit strong time-dependence and can be traversed by causal curves from past null infinity only if such curves depart at sufficiently early times.

Below, we review the general framework of \cite{Fu:2018oaq} and apply it to the asymptotically flat wormholes of interest here. As in \cite{Fu:2018oaq}, we define traversable wormholes to be the set of curves that can witness non-trivial topology while escaping out to infinity -- e.g. causal curves that cannot be deformed, while remaining causal, to lie in the boundary of the spacetime. To construct our relevant classical geometry, we start with a spacetime $\tilde M$ with a bifurcate Killing horizon and one asymptotic region on each side of the horizon; $\tilde M$ can be thought of as an {\it almost} traversable wormhole with two asymptotic regions. If this spacetime admits a freely-acting $\mathbb{Z}_2$ isometry $J$ exchanging the right and left asymptotic regions and preserving the time orientation, then the quotient $M = \tilde M/J$ describes an {\it almost} traversable wormhole with a single asymptotic region. While in principle, a small perturbation of either geometry could render the wormholes traversable, for $\tilde M$ the horizon generating Killing field forces the null stress-energy of any perturbation respecting this symmetry to be zero. In $M$, however, this Killing symmetry is broken by the quotient by $J$, which necessarily maps the horizon-generating Killing field $\xi$ to $-\xi$, since it identifies the right and left regions while preserving the time-orientation of the spacetime. This allows small perturbations to render the wormhole in the quotient space traversable. 

\begin{figure}
\begin{center}
\begin{subfigure}{.49\linewidth}
\centering
\includegraphics[height=4cm]{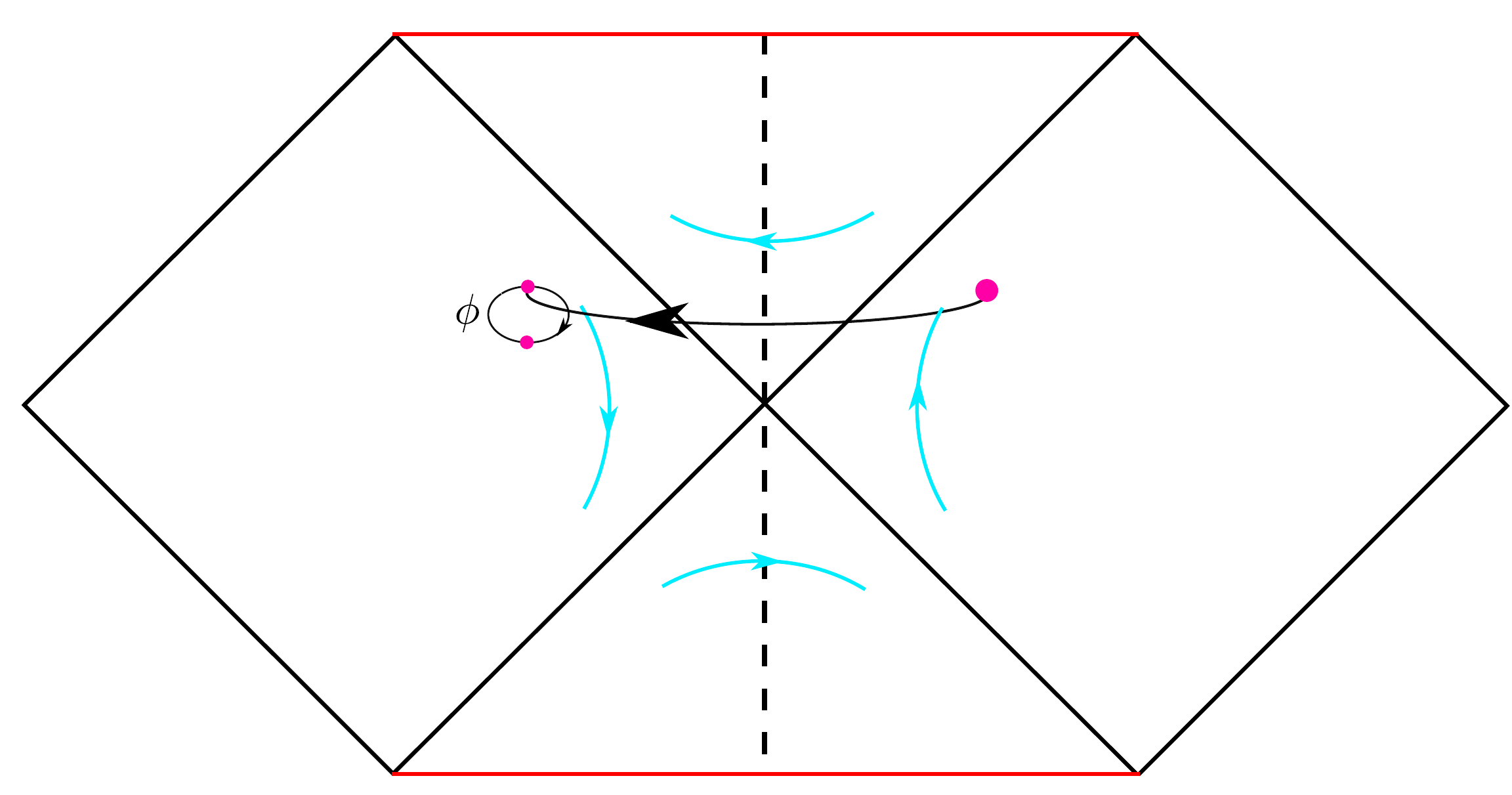}
\end{subfigure}
\begin{subfigure}{.49\linewidth}
\centering
\includegraphics[height=4cm]{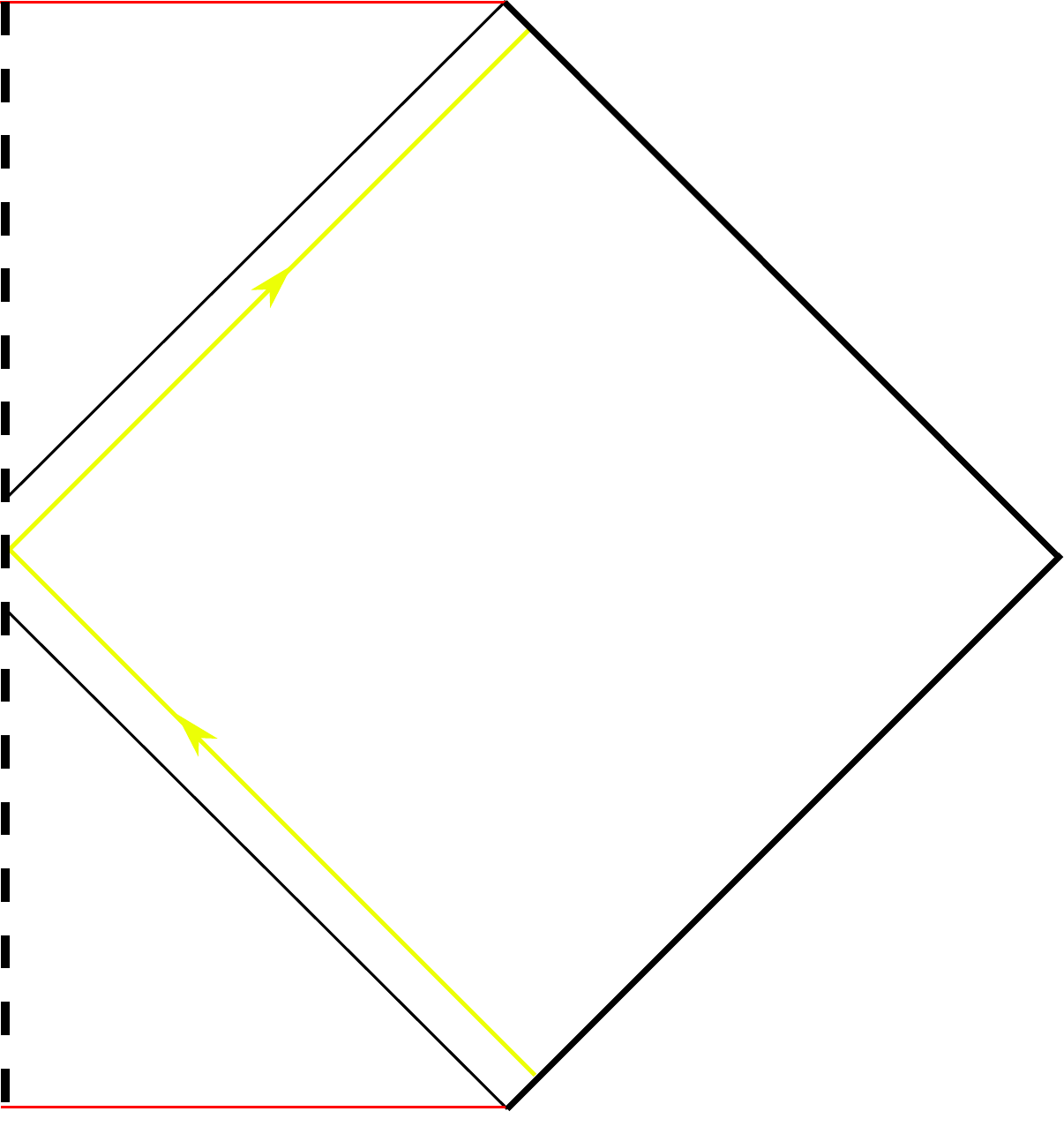}
\end{subfigure}
\end{center}
\caption{(Left) The $\mathbb{RP}^3$ geon is a $\mathbb{Z}_2$ quotient of the maximally-extended Schwarzschild black hole. The quotient acts on the above conformal diagram by reflection across the dashed line, and simultaneously acts as the antipodal map on the suppressed $S^1$. This action maps the Killing field $\xi^a$ to $-\xi^a$, and so the geon quotient lacks a globally defined time translation Killing field. In particular, the dashed line is orthogonal to preferred spacelike surfaces of vanishing extrinsic curvature that one may call $t = 0$. (Right) A small perturbation of maximally-extended Schwarzschild renders the $\mathbb{Z}_2$ quotient wormhole traversable. This results in a causal curve running between past and future null infinity that is not deformable to the boundary.}
\label{fig:model}
\end{figure}

\begin{figure}[t]
\begin{center}
\includegraphics[width=0.5\textwidth]{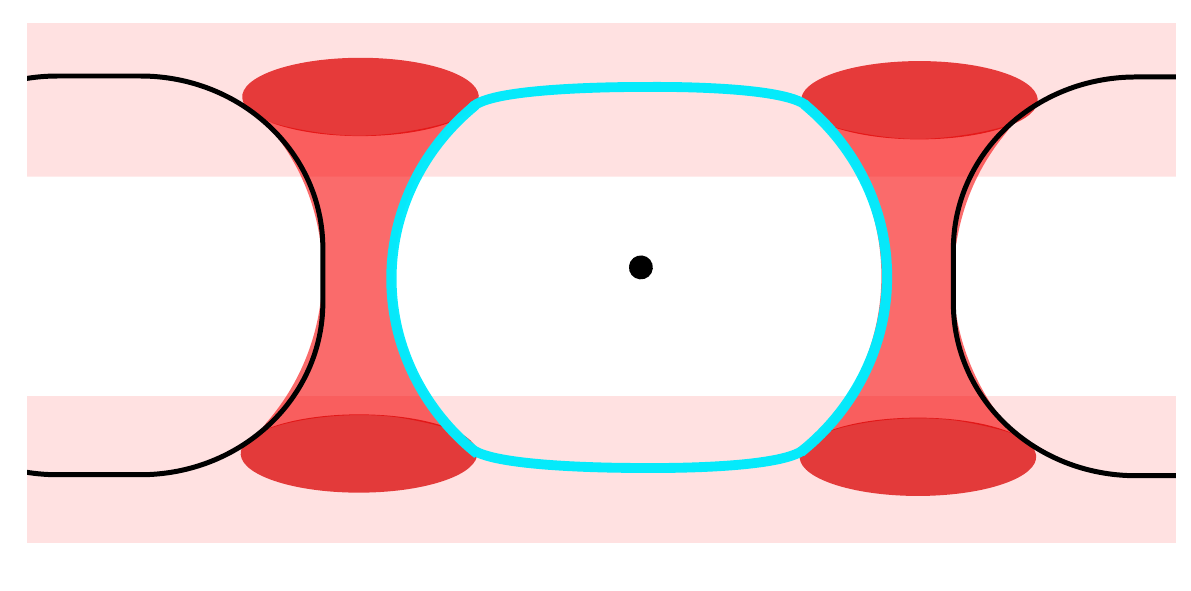}
\end{center}

\caption{A moment of time symmetry in our covering space $\tilde M$.  Each of the two asymptotic regions contains a pair of  black holes held apart by cosmic strings that stretch to infinity. A pair of strings thread the two wormhole throats and return to infinity in the second asymptotic region.  The ${\mathbb Z}_2$ isometry used to construct $M = \tilde M/{\mathbb Z}_2$ acts as a $\pi$ rotation about the non-physical point indicated by the dot at the center of the right figure.  As a result, a moment of time-symmetry for $M$ takes the form shown in figure \ref{fig:wormhole} and contains two wormhole mouths in a single asymptotic region.}
        \label{fig:BachWeyl}
\end{figure}

The simplest examples of such quotient wormholes are like the $\mathbb{RP}^3$ geon \cite{Misner:1957mt,GiuliniPhD,Friedman:1993ty} shown in figure \ref{fig:model}. Though the quotient $M$ then contains only a single wormhole mouth, it nevertheless admits causal curves $\gamma$ that are not deformable to the spacetime boundary. Similar spacetimes with asymptotically AdS$_3$ (or AdS$_3 \times X$) boundary conditions and their back-reaction from quantum scalar fields were explored in detail in \cite{Fu:2018oaq}, and similar back-reaction from bulk fermions will be explored in \cite{Sean}. Here, we will instead study the more sophisticated case where the covering space $\tilde M$ contains a pair of maximally-extended black holes, as in figure \ref{fig:BachWeyl}, so that the quotient $M$ takes the form depicted in figure \ref{fig:wormhole}. Since the particular solution $M$ studied below involves two cosmic strings, one stretching to infinity and the other compact, we require three cosmic strings in the covering space $\tilde M$. The compact cosmic string in $M$ lifts to a single longer compact cosmic string in $\tilde M$, while the string stretching to infinity in $M$ lifts to a pair of disconnected strings in $\tilde M$.

Once we have formed our classical backgrounds, it remains to understand the back-reaction from quantum fields sitting on the spacetime. As explained in \cite{Fu:2018oaq}, if quantum fields on $\tilde M$ have a well-defined Hartle-Hawking state, there will be a corresponding Hartle-Hawking-like state on $M$. This state is defined by the path integral over the appropriate quotient of the Euclidean geometry of $\tilde M$. For linear fields this state can also be constructed by applying the method of images to the Hartle-Hawking state on $\tilde M$. In particular, we can use the method of images to calculate expectation values of the stress tensor of quantum fields in their Hartle-Hawking state $\langle T_{kk}\rangle_{M} = \langle T_{ab}k^a k^b \rangle_{M} $ along affinely parameterized generators $k^a$ of the background spacetime's horizon. Because the stress tensor is a quadratic composite operator, the method of images implies that $ \langle T_{kk}\rangle_{M}$ in our $\mathbb{Z}_2$ quotient space can be written as 4 terms in our covering space. Two of these are just $\langle T_{kk}\rangle_{\tilde M}$ in the Hartle-Hawking state on $\tilde M$ which are forced to vanish by the Killing symmetry. The remaining terms involve two-point functions evaluated at some point $x$ on one horizon in the covering space $\tilde M$ and the image point $Jx$ under the isometry $J$, located on the other horizon. Since $J$ acts freely and the quotient $\tilde M/J$ contains no closed causal curves, the points $x$ and $Jx$ are spacelike separated and this two-point function is finite. The spacelike separation of $x$ and $Jx$ also guarantees that the two cross-terms coincide.

As mentioned above, though the Killing symmetry of the covering space forces $\langle T_{kk}\rangle_{\tilde M} =0$, breaking this symmetry by quotienting now allows non-zero $\langle T_{kk}\rangle_M$. More powerfully, the actual expression for $\langle T_{kk}\rangle_M$ depends on whether the quantum field is periodic or anti-periodic around the non-contractible cycle created by the quotient, and the two choices differ only by an overall sign. Because the effect of back-reaction on traversability is governed by the integral $\int \langle T_{kk}\rangle_M d\lambda$ (with respect to an affine parameter $\lambda$ over the generators of the horizon), barring surprising cancellations the above choice of sign allows us to tune the boundary conditions of our fields to render the wormhole traversable. For bosons, this sign tends to correspond to periodic boundary conditions in accord with the famous negative Casimir energy of periodic bosons on $S^1 \times {\mathbb R}^{d-1}$.

Returning to the asymptotically flat wormholes of interest here, we consider the stress-energy provided by the cosmic strings. Since the relevant null vectors $k$ are tangent to the cosmic string worldsheets and the classical cosmic string stress tensor is proportional to the induced metric, classically the cosmic strings do not contribute to $\langle T_{kk}\rangle_M$. Quantum fluctuations in the location of the string will contribute, however. We will model such fluctuations as 1+1 dimensional massless free scalar fields. Since correlators of 1+1 quantum fields diverge only logarithmically at short distance, it is easy to find a regime where these fluctuations remain small when compared with any classical scale (at least when the fluctuations are averaged over any classical time or distance scale), and our free field approximation is valid in such regimes. For the string stretching to infinity in $M$, the points $x, Jx$ in the covering space $\tilde M$ will lie on two distinct non-compact strings. Since fluctuations on two different strings are uncorrelated, the cross-terms in $\int \langle T_{kk}\rangle_M d\lambda$ will vanish. The contributions from quantum fluctuations of the compact string are non-zero, however, and studied below using a conformal transformation associated with the method-of-images construction described above.

At finite temperatures, we find that our wormholes become traversable for test signals and that, when the mouths are separated by a large distance $d$, well-timed signals require only the relatively short time $t_{\text{min\ transit}} = d + logs$ (in units with the speed of light $c$ set to 1) to traverse the wormhole, where $logs$ denote terms logarithmic in $d$ and in the black hole parameters. In particular, the transit time is shorter than for MPP wormholes by more than a factor of 2 and, as discussed in section \ref{sec:disc}, for $d\rightarrow \infty$ the higher-dimensional analogue of this result would approach the minimum transit time consistent with the above-mentioned prohibition against wormholes providing the fastest causal curves between distant points \cite{Wall:2013uza,Gao:2016bin}. However, in such cases traversability is exponentially fragile, and can be destroyed by exponentially small perturbations.

One should note that our background spacetime is unstable, as small perturbations will cause the black holes to either fall towards each other or fly apart. However, the time scale for the black holes to merge is $\sim d^{3/2}$, so the solutions are long lived compared to the transit time. Furthermore, one could engineer more complicated stable configurations using additional cosmic strings and anchoring to some stable structure at a finite distance (e.g., a large stable spherical shell surrounding both wormhole mouths) instead of running the strings off to infinity.  It will be clear below that the results for such more complicated models will be essentially the same.

Additionally, as in \cite{Maldacena:2018lmt,Maldacena:2018gjk,Fu:2018oaq}, we take particular interest in studying the extremal limit of our classical backgrounds. In \cite{Caceres:2018ehr,Fu:2018oaq} it was  shown that this limit gives large back-reaction for rotating BTZ, and we see here that this limit also gives large back-reaction for $d=4$ Reissner-Nordstr\"om black holes. On general grounds\footnote{We thank Zhenbin Yang for explaining this point.} this feature is related to the fact that far in the throat of a nearly extremal spherically-symmetric black hole, the size of the spheres is approximately constant, and thus one can approximately Kaluza-Klein reduce the dynamics to two-dimensional gravity. However, the Einstein-Hilbert action $\int \sqrt{g} R$ becomes a topological invariant in two dimensions, and does not contribute to the equations of motion, modeling the higher-dimensional case in the infinite coupling limit $G_N \rightarrow \infty$.  Thus, in the extremal limit,  the effective coupling diverges. Though our perturbation theory breaks down when the back reaction becomes large, we take the divergence as an indication that a full, non-perturbative calculation would reveal traversable wormholes that remain open for all time. 

In Section \ref{sec:simplemodel}, we compute
$\int \langle T_{kk}\rangle_M d\lambda$ from the Hartle-Hawking state quantum fluctuations of the cosmic strings. We then compute the back-reaction on our geometry and the resulting degree of traversability in section \ref{sec:grav}, and conclude with some brief remarks in section \ref{sec:disc}. As a contrasting side-note and because it provides an exactly solvable model for scalar fields of arbitrary mass,  we also compute  effects for what one may call a cosmological wormhole $dS_d/\mathbb{Z}_2$ in appendix \ref{sec:dS} where the back-reaction has the opposite sign so that negative energy from quantum fields in fact makes the wormhole harder to traverse.

\section{Stress-energy on the horizon}
\label{sec:simplemodel}

The introduction outlined a simple background spacetime $M$ with a wormhole whose mouths are held apart by cosmic strings.  This wormhole is not traversable, but is almost so  and will be rendered traversable by the back-reaction of quantum fields.  As noted in \cite{Fu:2018oaq}, the 2-fold covering space $\tilde M$ of this background is a charged version of the analytic extension behind the horizon \cite{Israel1964} of solutions found by Bach and Weyl in 1922 \cite{Bach1922}. For our case where the black holes have identical mass and opposite charge, an explicit form for this solution was found in \cite{Emparan:2001bb} based on the implicit solutions in \cite{Manko94}; see also \cite{Bonnor,Chandrasekhar:1989ds,Davidson:1994df,Emparan:1999au} for the simpler extreme case. Additionally, we wrap a compact cosmic string through both wormholes mouths, and our goal here is to understand any additional contributions to $\langle T_{kk}\rangle_{M}$ associated with its fluctuations. As discussed above, the contributions from any strings stretching to infinity will all vanish and we ignore contributions from bulk fields. We will not need the full details of the covering space (which can be found in the above references), as we will instead focus on ranges of parameters where the analysis simplifies. 

We will take the tension $\mu$ of the strings to be large compared with the length scale $r_0$ set by the black holes, $\mu r_0^2 \gg 1$, but we take Newton's constant $G_N$ even smaller ($G_N \mu \ll 1$) so that the conical deficit associated with the strings can be neglected. The first condition allows us to linearize the fluctuations, while the second means that we can ignore local effects of the strings on the geometry\footnote{Due to the logarithmic divergences of 1+1 field theories noted above, the first condition should really be $\mu r_0^2 \gg \ln n$ where $n$ is a parameter set both by the background spacetime and the manner in which fluctuations are averaged as described in detail below.}.  Note that the tension of strings stretching to infinity must be at least somewhat larger than that of the compact string in order to keep the black holes from coalescing. To suppress quantum fluctuations in $T_{kk}$, and also to justify neglecting the effects of bulk  Maxwell fields and linearized gravitons, we can replace each string in figure \ref{fig:wormhole} with $N$ strings so long as $\mu r_0^2 \gg 1$ for each string and $N G_N \mu \ll 1$.  As in \cite{Maldacena:2018gjk}, this will be necessary to render our semi-classical treatment in terms of expectation values valid.

To complete our specification of parameters, we further fix any measure of the distance $d$ between the mouths of our wormhole and take the limit where $d$ is much larger than the
the radius $r_+ = \sqrt{A/4\pi}$ of the black hole horizon. In this limit, the covering spacetime $\tilde M$ can be divided into three overlapping regions:  the region near the first black hole where the influence of the second can be treated as a small perturbation, the corresponding region for the second black hole, and the region in between where both black holes cause only small perturbations from flat space.  In each region, it is possible to systematically improve the approximation order by order in perturbation theory, but we work in the leading approximation below.

As noted in the introduction, it is possible to compute $\langle T_{kk} \rangle_{M}$ using the method of images starting in the covering space. However, since we are modeling fluctuations of the cosmic string as massless 1+1 free fields, they define a 1+1 dimensional conformal field theory. It is thus natural to instead compute $\langle T_{kk} \rangle_{M}$ by finding a conformal map from the 1+1 spacetime $M_{cs}$ induced on the worldsheet of the compact cosmic string to a piece of the cylinder of circumference $2\pi$, and which simultaneously maps the cosmic string Hartle-Hawking state to the cylinder vacuum.
Inverting the transformation will then determine our $\langle T_{kk} \rangle_{M}$ in terms of the known $\langle T_{ab} \rangle$ on the cylinder and the stress tensor Weyl anomaly associated with this conformal map. Such a map must exist since both the cosmic string Hartle-Hawking state and the cylinder vacuum can be constructed as path integrals over the respective spacetimes. Though we do not explicitly use the method of images to calculate $\langle T_{kk} \rangle_{M}$, we will still find the quotient construction to be of great use in finding this conformal map.  Below, we use $M_{cs} (\tilde M_{cs})$ to denote the 1+1 spacetimes induced on the compact cosmic string by $M (\tilde M)$, with $M_{cs} = \tilde M_{cs}/{\mathbb Z}_2$.

We first construct a conformal transformation relating $\tilde M_{cs}$ to a piece of a cylinder, and which maps the Hartle-Hawking state on $\tilde M_{cs}$ to the cylinder vacuum. The Killing symmetry of $\tilde M_{cs}$ means that its Hartle-Hawking state may be characterized as the unique Hadamard state invariant under the symmetry.  As a result, the pull-back of the cylinder vacuum under our conformal map will be the Hartle-Hawking state so long as the Killing symmetry of $\tilde M_{cs}$ maps to a symmetry of the cylinder vacuum.  Choosing locally-Minkowski coordinates $\phi, \tau$ on the cylinder (or equivalently null coordinates $u_c=\tau-\phi$, and $v_c=\tau+\phi$), we take this symmetry to be the 1-parameter subgroup of the vacuum-preserving SO(2,1) symmetry that acts like a boost near the origin $\phi = \tau = 0$ (or $u_c = v_c = 0$) and at appropriate other points that form a periodic array on the cylinder. For convenience after we take the $\mathbb{Z}_2$ quotient, we choose the unusual convention that here $\phi$ be periodic with period $4\pi$ on the cylinder conformal to $\tilde M$. 

\begin{figure}[t]
        \begin{center}
                \includegraphics[width=0.7\textwidth]{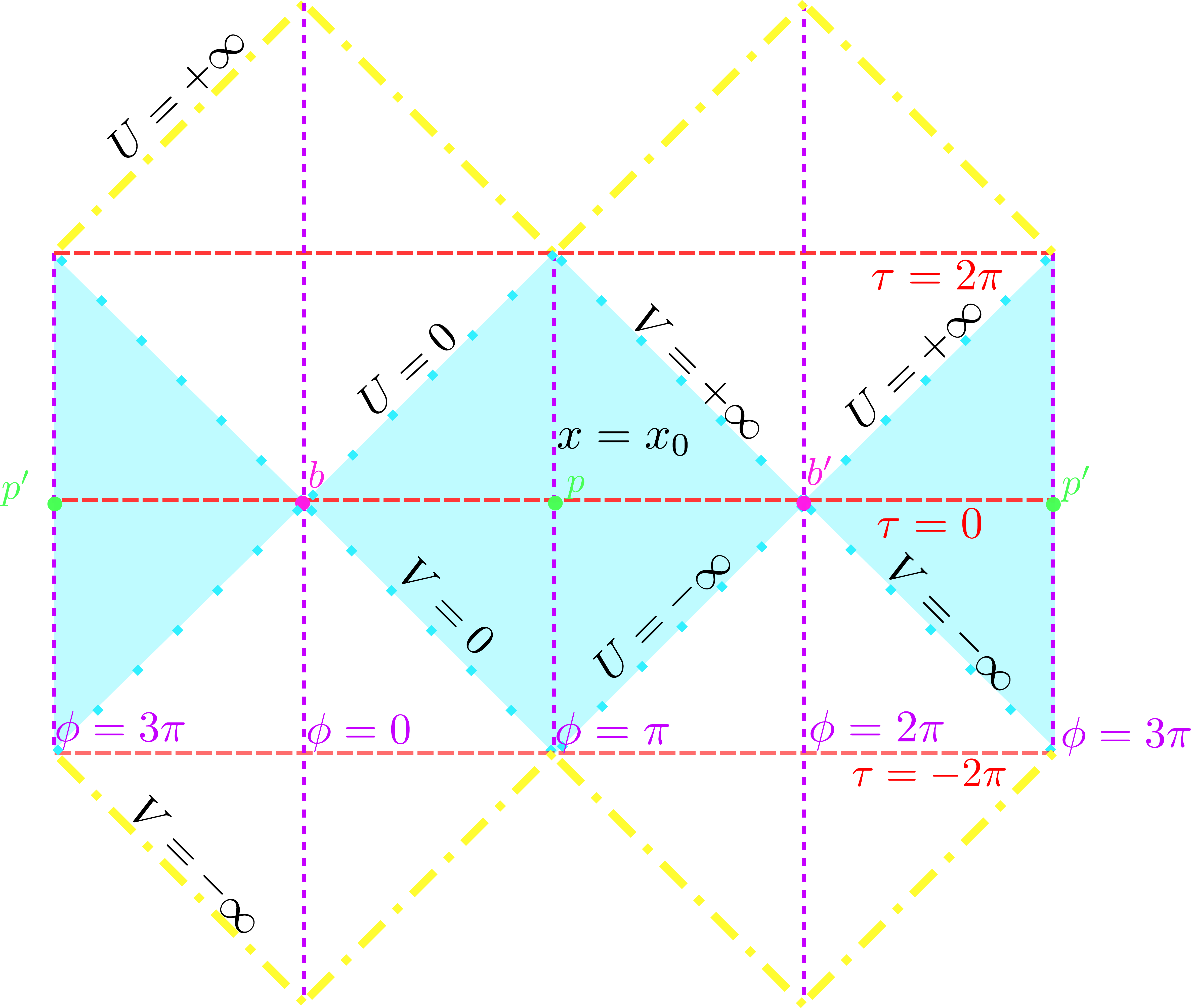}
                        \end{center}
	\caption{A conformal diagram of the two-fold cover ${\tilde M}_{cs}$ of the background spacetime $M_{cs}$ induced on our compact cosmic strings shown in the cylinder conformal frame.  Wormholes in this spacetime are not traversable, but are almost so and will be rendered traversable by the back-reaction of quantum fields.  The left and right edges of the diagram are to be identified because of the periodicity of $\phi$. Here we have truncated the spacetime at the black hole inner horizons (shown as dotted yellow lines) due to the expected instability of such horizons \cite{McNamara}. Blue dots mark the outer horizons, and the shaded regions are the static patches. The diagram is adapted to the cylinder coordinates $\phi, \tau$ as indicated by the horizontal dashed lines showing $\tau = -2\pi, 0, \pi$ and the vertical dashed lines showing $\phi = 0 , \pi, 2\pi, 3\pi$. (recall that $\phi$ has period $4\pi$ on ${\tilde M}_{cs}$).  The spacetimes has two bifurcation surfaces (points) $b, b'$ at $(\phi, \tau) = (0,0)$ and $(\phi, \tau) = (2\pi,0)$.  The points $p,p'$ at  $(\phi, \tau) = (\pi,0)$ and $(\phi, \tau) = (3\pi,0)$ are also marked, as are the right-moving null lines $U= - \infty, 0, +\infty$ and the left-moving null lines $V = -\infty, 0, +\infty$.}
        \label{fig:Mtildecs}
\end{figure}

In a static patch of $\tilde M_{cs}$ (see figure \ref{fig:Mtildecs}) the metric can be written in the form
\begin{equation}
\label{eq:smoothed}
ds^2_{\text{static}} = - f dt^2 + f^{-1} dx^2,
\end{equation}
with $f =f(x)$ and where $x$ ranges over $[0,2x_0]$. For later use, we also note that in the limit where the separation $d$ between the black holes satisfies $d \gg r_+$, there is a sphere of approximate symmetry passing through $x_0$ around either black hole, with radius \begin{equation}
\label{eq:rx0}
r(x_0) = x_0 + O(r_+) = d/2  + O({r_+}),
\end{equation}
where since the symmetry is only approximate the $O(r_+)$ terms depend on precisely how this sphere is defined\footnote{The second $O(r_+)$ term similarly depends on the precise definition of the separation $d$.}.

To map to the cylinder, we want $t,\phi$ such that
\begin{equation}
\label{eq:conf}
ds^2_{\text{static}} = \Omega^2 (-d\tau^2 + d \phi^2) = -\Omega^2 du_c dv_c
\end{equation}
for an appropriate conformal factor $\Omega$. Two such patches will be required to wrap around a piece of a cylinder conformal to $\tilde M_{cs}$ (see figure \ref{fig:Mtildecs}).  Therefore, in our static patch, $\phi$ ranges over $[0,2\pi]$.  This patch is thus precisely the part of the cylinder with $u_c \in [-2 \pi,0]$, $v_c \in [0, 2 \pi]$.

Before constructing the conformal map, we introduce Kruskal coordinates $U,V$ on $\tilde M_{cs}$. We build these in the usual way by first introducing the tortoise coordinate
\begin{equation}
x_* =x_{*0} +\int _{x_0}^{x}\frac{1}{f}dx,
\end{equation}
where we will fix the arbitrary parameter $x_{*0}$ below.  We then
define $u = t - x_*$, $v=t + x_*$, and finally 
\begin{equation}
U=-\kappa_+^{-1}e^{-\kappa_+u},\text{ }V=\kappa_+^{-1}e^{\kappa _+v},
\label{eq:kruskal}
\end{equation}
where $\kappa_+$ is the surface gravity of the black hole's outer horizon.  In these coordinates, metric becomes
\begin{equation}
\label{eq:simplemodel}
ds^2_{\text{static}}= {f}\left(-dt^2+dx_*^2\right) = \frac{f}{\kappa_+^2 UV}dUdV.
\end{equation}
Here the region $0 < x < 2x_0$ is mapped to $-\infty < x_* < + \infty$ and thus to $U\in(-\infty, 0)$, $V\in (0, \infty)$.  However, the form on the right-hand-side can be analytically continued to all points where both $U$ and $V$ are defined.  Comparing \eqref{eq:conf} with \eqref{eq:simplemodel} yields
\begin{equation}
\label{eq:omega}
\Omega^2 =- \frac{f}{\kappa_+^2 UV} \frac{dU}{du_c} \frac{dV}{dv_c}.
\end{equation}

Up to an arbitrary scale $L$, the symmetries determine the map to the cylinder to be
\begin{equation}
\label{eq:covertocyl}
U = L \tan(u_{c}/4),  \ \ \ V = L \tan(v_{c}/4).
\end{equation}
In particular, a map of this form also gives the standard conformal transformation relating the cylinder to 1+1 Minkowski space.  Here, we can fix $L$ by using the fact that we want the $\mathbb{Z}_2$ symmetry exchanging the black holes (and corresponding to a $\pi$ rotation about the non-physical point marked in figure \ref{fig:BachWeyl}) to correspond to half a rotation of the cylinder, which here is $\phi \rightarrow \phi + 2 \pi$; see again figure \ref{fig:Mtildecs}. Note that if the $\mathbb{Z}_2$ symmetry on $\tilde M_{cs}$ maps the null ray $U_1$ to the null ray $U_2$, the Killing symmetry requires that the $\mathbb{Z}_2$ symmetry map $\lambda U_1$ to $U_2/\lambda$. Using \eqref{eq:covertocyl} in addition, we see that  the action of this $\mathbb{Z}_2$ is
\begin{equation}
U \rightarrow - L^2/U, \ \ \ V \rightarrow - L^2/V.
\end{equation}
We can fix $L$ by using time-reversal symmetry to note that the null lines through the point $p$ (with coordinates $t=0, x=x_0$) and its image $p'$ under this
$\mathbb{Z}_2$ have $U=\pm \kappa_+^{-1}e^{\kappa_+ x_{*0}}$ and have $u_c=-\pi,\pi$, so that $L=\kappa_+^{-1}e^{\kappa_+ x_{*0}}$. 

Having constructed the conformal map from $\tilde M_{cs}$ to the cylinder with period $4\pi$, it is now straightforward to take the $\mathbb{Z}_2$ quotient and use this same conformal map to relate $M_{cs} = {\tilde M}_{cs}$ to the cylinder with period $2\pi$.  Here it is of course critical that we ensured  the original conformal map took the $\mathbb{Z}_2$ action on ${\tilde M}_{cs}$ to the action $\phi \rightarrow \phi +2\pi$ on the cylinder.  Since the original map related the Hartle-Hawking state on $\tilde M_{cs}$ to the $4\pi$ cylinder vacuum, the method of images guarantees that it also relates the Hartle-Hawking state on $M_{cs} = {\tilde M}_{cs}$  to the vacuum on the standard cylinder of period $2\pi$ as desired.

This map can now be used to compute the integrated null stress tensor $\int T_{kk} d\lambda$ in the Hartle-Hawking state of $M_{cs}$. To simplify this, we introduce rescaled Kruskal coordinates
\begin{equation}
\bar U = \frac{U}{L \kappa_+} = e^{-\kappa_+ x_{*0}} U, \ \ \ \bar V = \frac{V}{L \kappa_+} = e^{-\kappa_+ x_{*0}} V.
\end{equation}
Since any Kruskal coordinate is an affine parameter on the horizon, we compute
\begin{equation}
\label{eq:bUbUint}
\int T_{\bar U \bar U} d\bar U = \int \frac{d u_c}{d \bar U} T_{u_c u_c} du_c,
\end{equation}
where the integral is performed over the horizon $V = \bar V =0$.
Due to the Weyl anomaly (see e.g. \cite{DiFrancesco:1997nk} as translated to standard Lorentz signature conventions by \cite{Fischetti:2012ps}), for any null vector $\hat k$ the component $T_{\hat k \hat k}$ is related to the associated components of the cylinder vacuum stress tensor $T^{\text{cyl}}_{\hat k\hat k} = \hat k^a \hat k^b T^{\text{cyl}}_{ab}$ by
\begin{equation}
\label{eq:TWeyl}
T_{\hat k \hat k}=T^{\text{cyl}}_{\hat k \hat k}+\frac{c}{12\pi }\left\{\nabla _{\hat k}\nabla _{\hat k}\left(\ln \Omega \right)-\left[\nabla _{{\hat k}}\left(\ln \Omega \right)\right]^2\right\},
\end{equation}
where $\nabla_{\hat k} = \hat k^a \nabla_a$, the covariant derivative is defined by the metric $ds^2_{\rm cyl} = - du_c dv_c$ on the standard cylinder, and we have used the fact that $\hat k^a$ is null.

$T_{u_c u_c}$ is then found by setting $\hat k^a  \partial_a = \partial_{u_c}$ in \eqref{eq:TWeyl}.  In particular, in the standard cylinder vacuum we have
\begin{equation}
T^{\text{cyl}}_{ab}=\rho (dt_{\text{cyl}})_a(dt_{\text{cyl}})_b+\rho (d\phi _{\text{cyl}})_a(d\phi _{\text{cyl}})_b,
\end{equation}
where $\rho =-\frac{c}{24\pi }$
and $c$ is the CFT central charge.  Thus
\begin{equation}
\label{eq:cylTucuc}
T^{\text{cyl}}_{u_c u_c} = -\frac{c}{48\pi}.
\end{equation}
Since our bulk spacetime has $3+1$ dimensions, there are two transverse polarizations for oscillations of the string. For our $N$ compact cosmic strings, this yields $c=2N$.

To compute the remaining terms in \eqref{eq:TWeyl} it is useful to observe that $\nabla_{u_c} = \partial_{u_c}$ since $u_c$ and $v_c$ are affine on the cylinder, and that
\begin{equation}
\label{eq:simplify}
\partial^2_{u_c} \left(\ln \Omega \right)-\left[\partial_{{u_c}}\left(\ln \Omega \right)\right]^2
= - \Omega \partial^2_{u_c} \left(\Omega^{-1} \right).
\end{equation}
 Since we only need to compute $\langle T_{kk}\rangle_M$ on the horizon $V = \bar V =0$, it is useful to recall that $\frac{f}{UV} = 2 g_{UV}$ is constant over the horizon as required to make $U$ affine there.  The factor $\frac{dV}{dv_c}$ is also constant on lines of constant $V$.  The only $u_c$-dependent factor in \eqref{eq:omega} is thus
\begin{equation}
\frac{dU}{du_c}  = \frac{L}{4 \cos^2(u_c/4)},
\end{equation}
so $\Omega^{-1} \propto  \cos^2(u_c/4)$ and
\begin{equation}
\label{eq:2ndOmega}
 - \Omega \partial^2_{u_c} \left(\Omega^{-1} \right) = \frac{1}{16}.
\end{equation}
Combining \eqref{eq:TWeyl}, \eqref{eq:cylTucuc}, \eqref{eq:simplify}, \eqref{eq:omega}, and \eqref{eq:2ndOmega} yields
\begin{equation}
\label{eq:Tucuc}
T_{u_c u_c} = - \frac{c}{64 \pi},
\end{equation}
so that \eqref{eq:bUbUint} yields
\begin{equation}
\label{eq:bubuint2}
\int T_{\bar U \bar U} d\bar U = \int_0^{2 \pi} 4 \kappa_+ \cos^2(u_c/4) \left(  - \frac{c}{64 \pi} \right) du_c = - \frac{c \kappa_+}{16},
\end{equation}
and thus
\begin{equation}
\label{eq:uuint}
\int T_{U U} d U = \int \frac{d\bar U}{dU} T_{\bar U \bar U} d\bar U =  e^{-\kappa_+ x_{*0}}\int T_{\bar U \bar U} d\bar U = - e^{-\kappa_+ x_{*0}} \frac{c \kappa_+}{16}.
\end{equation}

Finally, it remains to choose the constant $x_{*0}$.  Since this was an arbitrary constant that entered only through the definition of a coordinate, physical results like the back-reaction of quantum fields on the geometry cannot depend on its value.  But the value of $\int T_{U U} d U$ does depend on the normalization of $U$, and it useful to make a choice that illustrates the relevant physics already at this stage.  Recall that the null ray through the point $x_0$ has $U = - \kappa_+^{-1} e^{\kappa_+ x_{*0}}$ at $t=0$, and this ray passes through an approximate sphere around either black hole of radius $r(x_0) = d/2  + O({r_+})$; see \eqref{eq:rx0}.  Standard dimensionful Reissner-Nordstr\"om Kruskal coordinates $U = -\kappa_+^{-1} e^{\kappa_+(r_*-t)}$ are defined using a tortoise coordinate with $r_* = r + O(\ln \frac{r}{r_+})$, and so we choose
\begin{equation}
\label{eq:xs0}
x_{*0} = d/2 +  O(\ln \frac{d}{r_+}),
\end{equation}
which yields $U(t=0, x_* = x_{*0}) = - \kappa_+^{-1}e^{\kappa_+\left({d/2} + O(\ln \frac{d}{r_+})\right)} $.

At finite $\kappa_+ > 0$, the stress-energy is thus exponentially small in the black hole separation $d$.   So while the negative sign in \eqref{eq:uuint} should make the wormhole at least formally traversable, this result will be exponentially sensitive to further perturbations -- including that from any signal sent through the wormhole.  We will return to such issues in section \ref{sec:disc} after carefully computing the back-reaction from \eqref{eq:uuint} in section \ref{sec:grav}.  For now, we simply note that this contrasts sharply with the expectation of a Casimir energy of order $1/d$ for large $d$ at fixed $\kappa_+$.  The difference is due in part to the fact that the integrated null energy \eqref{eq:uuint} differs from the conserved total energy of the quantum field by a factor of $\xi^U$, the null component of the Killing field which would appear in the latter but does not enter \eqref{eq:uuint}.  As a result, energy that falls across the horizon at late times is exponentially suppressed in \eqref{eq:uuint} relative to the conserved total energy. 

Before proceeding, we pause to note that the result \eqref{eq:Tucuc} could in fact have been predicted without calculation by combining the following observations. First, the fact that $U$ is affine along the horizon means that we could conformally map our physical spacetime to flat 1+1 Minkowski space using a conformal factor $\tilde \Omega$ that is constant on the horizon and which thus has no anomalous contribution to the associated null-null stress-energy.  Second, the standard conformal map from 1+1 Minkowski to the cylinder maps the Minkowski vacuum to the cylinder vacuum and thus has an anomaly that precisely cancels the cylinder stress tensor \eqref{eq:cylTucuc}. However, thirdly, our $\mathbb{Z}_2$ quotient introduces factors of 2 that scale the anomalous contribution by $1/4$, so that it will only partially cancel the cylinder stress-energy.  Thus \eqref{eq:Tucuc} is precisely $3/4$ of \eqref{eq:cylTucuc}.   The rest of the computations simply apply this rescaled version of the standard conformal map from the plane to the cylinder.  As a result, the final expression \eqref{eq:uuint} must in fact be the identical for any other 1+1 background with the same causal structure up to the choice of $x_{*0}$ that determines the overall scale of the effect.

\section{Back-reaction and Stability}
\label{sec:grav}

We are now ready to study first-order back-reaction from the quantum stress-energy \eqref{eq:uuint}, and in turn, study the traversiblity of our wormhole. 
We first orient ourselves to the appropriate geometry in section \ref{subsec:geom} before investigating the linearized Einstein equations in section \ref{subsec:lin}.

\subsection{Geometry and Geodesics}
\label{subsec:geom}

We are primarily interested in following a null geodesic through the throat of our wormhole.  As described above, when the wormhole mouths are far apart, the spacetime in the throat is approximately spherically symmetric and thus Reissner-Nordstr\"om up to small corrections.  At leading order in large $d$, it thus suffices to study perturbations to Reissner-Nordstr\"om sourced by the stress-energy \eqref{eq:uuint}, and in particular on our null geodesic.

We start with the Reissner-Nordstr\"om metric in its static form
\begin{equation}
\label{eq:rnmetric}
ds^2 = - f dt^2 + f^{-1}dr^2 + r^2 d\Omega^2,
\end{equation}
where $d\Omega^2$ is the metric on the unit two-sphere. As above, we introduce Kruskal coordinates: the standard tortoise coordinate is \begin{equation}
\label{eq:tortoise}
{{r}_{*}}=\int{\frac{1}{f}dr}=r+\frac{1}{2{{\kappa }_{+}}}\ln \frac{\left| r-{{r}_{+}} \right|}{{{r}_{+}}}-\frac{1}{2{{\kappa }_{-}}}\ln \frac{\left| r-{{r}_{-}} \right|}{{{r}_{-}}}
\end{equation}
where we have chosen the constant of integration such that $r_*=0$ at $r=0$. We then introduce $u = t - r_*$, $v = t + r_*$ and thus the dimensionful Kruskal coordinates 
\begin{equation}\label{eq:kruskal2}
U = \mp \kappa_+^{-1} e^{\mp\kappa_+ u}, \ \ \ V = \pm \kappa_+^{-1} e^{\pm\kappa_+ v},
\end{equation}
where the $(U,V)$ signs are $(-,+)$ before the geodesic enters the throat and are $(+, -)$ after it leaves.  The metric in these coordinates becomes
\begin{equation}
d{{s}^{2}}=2g_{UV}dUdV+{{r}^{2}}\left( d{{\theta }^{2}}+{{\sin }^{2}}\theta d{{\phi }^{2}} \right),
\end{equation}
where $r=r(UV)$ is implicitly defined by equations \eqref{eq:tortoise} and \eqref{eq:kruskal2}. As usual, we have
\begin{equation}\label{eq:RN}
g_{UV}=g_{UV}(UV)=\frac{f}{2\kappa_+^2UV}=-\frac{1}{2}\frac{{{r}_{+}}{{r}_{-}}}{{{r}^{2}}}{{\left( \frac{r-{{r}_{-}}}{{{r}_{-}}} \right)}^{1+{{\left( {{r}_{-}}/{{r}_{+}} \right)}^{2}}}}{{e}^{-2{{\kappa }_{+}}r}}.
\end{equation}
The null curve $V=0$ at constant angles on the $S^2$ is a geodesic in this background.  Following the standard treatment, we wish to understand how this geodesic is displaced under a general metric perturbation $h_{ab}$.  Integrating the geodesic equation gives
\begin{equation}
\label{eq:pertV}
V(U) = -(2 g_{UV}(V=0))^{-1} \int_{-\infty}^{U} \text{d}U h_{kk},
\end{equation}
where $h_{kk} = h_{ab} k^a k^b$ and where we have used the fact that $g_{UV}$ is constant along the unperturbed horizon at $V=0$. At $U= +\infty$ one thus finds
\begin{equation}
\label{DeltaV}
\Delta V = \frac{r_+}{r_-}{{\left( \frac{r_+-r_-}{r_-} \right)}^{-1-{{\left( {{r}_{-}}/{{r}_{+}} \right)}^{2}}}}{{e}^{2{{\kappa }_{+}}r_+}} \int_{-\infty}^{+\infty} \text{d}U h_{kk}.
\end{equation}
So long as this quantity is negative, the geodesic will emerge from the black hole and reach null infinity.  As in \cite{Gao:2000ga,Fu:2018oaq}, we will see in section \ref{subsec:lin} below that negative $h_{kk}$ follows from the negative $\langle T_{kk}\rangle$ found above in \eqref{eq:uuint}.

In addition to the binary question of traversability, we can also study the time-delay of this wormhole-traversing null geodesic relative to some standard. For reference purposes, let us consider a non-physical (particularly violating the generalized second law) ultrastatic ($g_{tt}= -1$) spacetime consisting at each time of two copies of Euclidean space, each with a ball of radius $r_+$ removed around the origin and with the two spheres glued together.  A null ray hitting one of these spheres in the first space then instantly teleported to an associated point in the other.  Note that Eddington-Finklelstein coordinates on such a space with the above conventions would have $v = \text{constant}$ for a radial null ray traveling from one asymptotic region to the other. As a result, if a null geodesic through our wormhole has $v_{\rm out} = v_{\rm in}$, it is as if the wormhole brought it instantaneously from one mouth to the other.  Conversely, with these conventions a null geodesic in Minkowski space that takes a time $d$ to travel the distance $d$ separating the mouths has $v_{\rm out} - v_{\rm in} =  d$.  As a result, the time delay relative to geodesics that propagate across the same separation in Minkowski space is\footnote{As always for $d=4$, in our actual background with black holes, propagation through the $1/r$ potential gives an additional logarithmic delay. However, it is still conventional to discuss time delay relative to comparable travel through Minkowski space.} $t_{\rm delay} = v_{\rm out} - v_{\rm in} - d$, so that it is natural to refer to $v_{\rm out} - v_{\rm in}$ as the transit time $t_{\rm transit}$ required for the signal to traverse the wormhole.

We should thus compute
\begin{equation}
\label{eq:39}
t_{\rm transit} = v_{\rm out} - v_{\rm in} = v(V(U=+\infty)) - v(V(U=-\infty))
\end{equation}
from \eqref{DeltaV}. Due to the exponential relationship between $v$ and $V$, \eqref{eq:39} is minimized for a geodesic starting at $V = -\Delta V/2$ and ending at $V = +\Delta V/2$ so that
\begin{equation}
\label{eq:tmd}
t_{\text{min transit}} = v_{\rm out}\left(- \frac{\Delta V}{2}\right) - v_{\rm in}\left(-\frac{\Delta V}{2}\right) = - \frac{2}{\kappa_+} \ln \left(-\kappa_+ \frac{\Delta V}{2} \right).
\end{equation}
From \eqref{eq:uuint} -- and the fact that we work in linear perturbation theory -- we thus expect to find $t_{\text{min transit}} =  2 x_{*0} + \text{logs} = d + \text{logs}$ so that the ratio $\frac{t_{\text{min transit}}}{d}$ to the transit time for a geodesic that does  not pass through the wormhole becomes $1$ in the limit of large $d$.  This expectation will be confirmed below.

\subsection{Back-reaction}
\label{subsec:lin}

We now study the metric perturbation $h_{ab}$ associated with the quantum stress-energy \eqref{eq:uuint}.  As noted above, at leading order in $d$ it suffices to perturb around the exact Reissner-Nordstr{\"o}m metric \eqref{eq:rnmetric}, and we consider a general perturbation. We will also need the Reissner-Nordstr\"om electromagnetic field, which in Kruskal coordinates takes the form
\begin{equation}
{{F}_{ab}}=-\frac{Q}{2{{\kappa }_{+}}}\left[ \frac{1}{V}{{\partial }_{U}}\left( \frac{1}{r} \right)+\frac{1}{U}{{\partial }_{V}}\left( \frac{1}{r} \right) \right]{{\left( dU \right)}_{a}}\wedge {{\left( dV \right)}_{b}}.
\end{equation}
One might expect that we also need to consider the spherically perturbed electromagnetic field $F_{ab} + \delta F_{ab}$.  However, because the electromagnetic stress tensor is quadratic in $F_{ab}$, and since the component $T_{UU}^{\text{(EM)}}$ vanishes by symmetry in the unperturbed background, it turns out that to first order one finds simply
\begin{equation}
\label{eq:TEM}
\delta T_{UU}^{\text{(EM)}}=-\frac{Q^2}{8\pi r^4}{h_{UU}} = -\frac{r_+r_-}{8\pi r^4} h_{UU}
\end{equation}
which is independent of $\delta F_{ab}$.  Therefore, on the horizon $V=0$, the $UU$ component of the linearized Einstein equations becomes
\begin{equation}
\begin{aligned}
\label{eq:linEE}
8\pi G T_{UU}^{\text{(scalar)}}=\frac{{{\kappa }_{+}}}{{{r}_{+}}}\left( 2h_{UU}+U\partial_{U}h_{UU} \right)-\frac{1}{2r_{+}^{2}}\partial_U^2\left(h_{\theta \theta}+\frac{1}{\sin^{2}\theta}h_{\phi \phi } \right)+\frac{1}{2r_{+}^{2}}\left[-\partial_{\theta }^2h_{UU} \right. & \\
\left. -\frac{1}{\sin^{2}\theta}\partial_\phi^2h_{UU}-\cot \theta\partial_{\theta}h_{UU}+2\cot \theta\partial_{U}h_{U\theta }+\frac{2}{\sin^{2}\theta}\partial_{U}\partial_{\phi }h_{U\phi }+2\partial_{U}\partial_{\theta }h_{U\theta} \right]. &
\end{aligned}
\end{equation}

We may then follow \cite{Fu:2018oaq} in integrating \eqref{eq:linEE} over $U$ at each point on the $S^2$ and applying asymptotically flat boundary conditions to find
\begin{equation}
\label{eq:inthT}
 8 \pi G \int \langle T_{kk}\rangle \text{d}U= \left(\frac{\kappa_+}{r_+}+\frac{1}{2r_+^2}(-\partial_\theta^2 - \frac{1}{\sin^2\theta}\partial_\phi^2 - \cot\theta \partial_\theta)\right)\int h_{UU} \text{d}U.
\end{equation}
Because we are interested in solving for the perturbation to the metric in terms of the stress tensor, we can invert this by finding an appropriate Green's function, $H(\Omega, \Omega')$ on $S^2$: 
\begin{equation}
\label{eq:hfromT}
\left( \int dU h_{kk} \right) (\Omega) = 8 \pi G \int d \Omega' H(\Omega, \Omega') \int dU \langle T_{kk} \rangle (\Omega').
\end{equation}
As usual, the general, explicit expression for $H(\Omega, \Omega')$ is rather cumbersome, but here it suffices to consider the response our compact cosmic string, which gives $ \langle T_{kk} \rangle (\Omega')$ proportional to a delta-function at a single point on the $S^2$, which we  take to be the north pole $\theta' =0$.  The remaining rotational symmetry then makes $H$ a function only of the polar angle $\theta$, reducing to the known Green's function for the Helmholtz equation \cite{Szmytkowski_2007}:  
\begin{equation}
\label{eq:GF}
H(\theta) = -\frac{r_+^2}{2  \sin (\pi \lambda)} P_\lambda (-\cos \theta)
\end{equation}
for $\lambda= -\frac{1}{2}\left(1+\sqrt{1-8\kappa_+r_+}\right)$, and $P_l(x)$ the Legendre polynomial, or equivalently,
\begin{equation}
\label{eq:GFFourier}
H = \sum_{j} Y_{m=0, j}(\Omega) H_{mj}, \ \ \ H_{mj} =\sqrt{\frac{2 j+1}{ 4\pi}} \frac{2r_+^2}{2\kappa_+ r_+  + j(j+1)}
\end{equation}
where $Y_{m=0, j}(\Omega)  = \sqrt{\frac{2j+1}{4\pi}}P_j(\cos \theta)$ are standard scalar spherical harmonics on $S^2$ with vanishing azimuthal quantum number.   As in \cite{Fu:2018oaq}, we find that the response $H_{mj}$ is largest at small $j$, and that this effect becomes very strong at small $\kappa_+$ in which case $H_{j=0}$ becomes very large.

Note that \eqref{eq:GF} is everywhere positive, and that it is largest at the north pole (where our compact cosmic string resides). The minimal transit time is thus experienced by the geodesic at $\theta=0$.  But for general $\theta$  \eqref{eq:uuint}, \eqref{eq:tmd}, \eqref{eq:hfromT}, and \eqref{eq:GF} yield

\begin{align}
\label{eq:mt}
t_{\text{min transit}}(\theta) &= 2x_{*0}  -4r_+ -\frac{2}{\kappa_+} \ln \left(\frac{\pi G c}{16}\frac{r_-}{r_+^3} \left(\frac{r_+-r_-}{r_-}\right)^{1-(r_-/r_+)^2} H(\theta)\right).
\end{align}

While $H(\theta)$ diverges for small theta, the divergence is only logartithmic. Since it also appears inside another log in \eqref{eq:mt}, the effect of this divergence is thus rather small and is also independent of $d$. Using \eqref{eq:xs0} thus gives $t_{\text{min transit}}(\theta)\approx d$ up to terms that grow no faster than logarithmically at large $d$.

\section{Discussion}
\label{sec:disc}

In the above work we studied the back-reaction from quantum fields in their Hartle-Hawking state on a simple classical wormhole solution of general relativity of the form shown in figure \ref{fig:wormhole}.  In the unperturbed solution, both of the wormhole mouths are black holes, and the wormhole interior collapses to a singularity.  In particular,
since the background respects the NEC, the background wormhole is non-traversable as predicted by topological censorship \cite{Friedman:1993ty,Galloway:1999bp}.  The solution of interest is a charged version of that first constructed by Bach and Weyl in 1922 \cite{Bach1922}, and contains cosmic strings which hold the two mouths of the wormhole apart at some separation $d$ and prevent them from coalescing.  Adding charge to the Bach-Weyl solution allows one to adjust the surface gravity $\kappa_+$ of the black holes. The solution is asymptotically flat apart from the fact that some of these cosmic strings stretch to infinity.  An explicit form for such solutions can be found in \cite{Emparan:2001bb} based on the implicit solutions in \cite{Manko94}; see also \cite{Bonnor,Chandrasekhar:1989ds,Davidson:1994df,Emparan:1999au} for the simpler extreme limit. 

While the wormhole is not traversable, it is infinitesimally close to being so in the sense that one can find two null rays separated by an arbitrarily small amount at $t=0$ such that one ray begins at past null infinity and enters one mouth of the wormhole while the other exits the other mouth and reaches future null infinity.    As a result, an arbitrarily small change in the metric generated by perturbative back-reaction from the stress-energy of quantum fields can render the wormhole traversable, at least for some period of time.  In  this work we computed the expected stress-energy associated with fluctuations in the locations of the of cosmic strings in their Hartle-Hawking state, as well as the first-order back-reaction of this stress-energy on the metric. When the number $N$ of such strings is sufficiently large, this expectation value should dominate over any fluctuations in this quantity, and also over contributions from bulk fields (e.g., from linearized gravitons) neglected in this work.  However, aside from greybody factors associated with the propagation of such fields into the wormhole throat, it is natural to expect contributions from bulk fields to be qualitatively similar to those found here for cosmic string fluctuations.

As expected on general grounds, here periodic boundary conditions give negative integrated null energy on the horizon. This defocuses null geodesics and slows the collapse of the wormhole, allowing properly chosen causal curves to traverse the wormhole and avoid the singularities.  In particular, these curves must begin their traversal of the wormholes at sufficiently early times.  Note that we have not computed the full back-reacted metric sourced by our quantum fields, but (following \cite{Gao:2016bin}) we have focused on showing that a particular class of causal curves can traverse the perturbed wormhole and on computing the time-advance that defines the associated transit times.  For contrast, appendix \ref{sec:dS} describes a `cosmological wormhole' in which the back-reaction of negative quantum stress-energy causes a time-{\it delay} instead of the above time-advance. As in \cite{Maldacena:2018lmt,Maldacena:2018gjk,Caceres:2018ehr,Fu:2018oaq}, the time advance in our asymptotically-flat case becomes large in the limit $\kappa_+ d \rightarrow 0$ where the background black holes become extremal.  Our perturbative description then breaks down but, at least at large $N$, it is natural to expect non-perturbative corrections to render the wormhole traversable for all time as in \cite{Maldacena:2018gjk}.

However, we can more concretely discuss the non-extremal case where perturbation theory is valid.  Although the integrated null energy on the horizon remains negative and proportional to $N$, it also becomes exponentially small in $\kappa_+ d$ (of order $e^{-\kappa_+ d/2}$).  The resulting traversability is thus extremely fragile, as an exponentially small positive-energy perturbation will negate this effect and prevent traversability\footnote{\label{foot:self} This includes possible perturbations associated with any signal one might attempt to send through the wormhole.  Now, a right-moving signal is sensitive to the back-reaction of left-moving stress-energy, and in a pure 1+1 massless theory, a right-moving signal will generate only right-moving stress-energy and so will not interfere with its own attempt to traverse a wormhole.  But more generally, right-moving signals will generate some amount of left-moving stress-energy as well. For example, in our model left- and right-moving oscillations of the cosmic strings are coupled via their interactions with the 4-dimensional bulk gravity.  However, in the covering space $\tilde M$ it is clear that any associated self-delay effect is independent of when the signal is sent into the black hole.  As a result, the integrated null stress-energy defined by any fixed affine parameter along the horizon must be exponentially small when a signal enters at early times.  As a result, to protect a weak signal entering the left mouth of the wormhole on $M$ from a strongly-blueshifted version of its own back-reaction it suffices to prevent the signal from sending perturbations into the {\it other} (right) mouth at early times.  Nevertheless, it would be interesting to study this back-reaction in detail as was done for GJW wormholes in \cite{Maldacena:2017axo,Caceres:2018ehr,Hirano:2019ugo,Freivogel:2019whb}, as this will place fundamental limits on the amount of information that can be transmitted.  We thank Eduardo Test\'{e} Lino for discussions on this point.}.

The exponentially small integrated null energy is due in part to the fact that we integrate affine stress-energy components $T_{UU}$ which are exponentially redshifted for energy that falls into the black hole at late times; i.e., because the Casimir-like energy located at $r\sim d/2$ at $t=0$ takes a time of order $d$ to fall into the black hole.  The effect of positive-energy perturbations is similarly suppressed at late times, so it is only necessary to be exponentially careful with our solution for a time of order $d$ after $t=0$.  At later times a more modest (though still significant) degree of care suffices to allow a signal to pass through, at least modulo the comments of footnote \ref{foot:self}.

This exponentially small expected stress-energy may also make one ask again about quantum fluctuations.  But as described in \cite{Gao:2016bin}, the fluctuations of integrated null stress-energy are exactly zero in the Hartle-Hawking state of our double-cover spacetime $\tilde M$.  While fluctuations on the quotient $M$ will be non-zero due to image terms much like those that give non-vanishing $T_{UU}$, they will again be exponentially suppressed\footnote{This assumes that we leave the system isolated for a time of order $d$, and in particular that we do not attempt to detect the signal before this time.  The response to sampling the system earlier would involve an integral of $T_{UU}$ supported on only part of the real line, in which case its fluctuations will not vanish even on the covering space $\tilde M$.  So such sampling could easily provide the exponentially small positive perturbation required to prevent traversability.}.  So the standard $\sqrt{N}$ suppression of fluctuations relative to the mean will suffice to protect traversability at only moderately large $N$.

Despite the small integrated stress-energy, the actual transit time through the wormhole is rather short.  In particular, for a wormhole with mouths separated by a distance $d$, we find $\frac{t_{\text{min\ transit}}}{d} \rightarrow 1$ as $d \rightarrow \infty$; see \eqref{eq:mt}.  As mentioned in the introduction, general arguments (and in particular the generalized second law) prohibit wormholes from providing the fastest causal curves between distant points \cite{Wall:2013uza,Gao:2016bin}. Naively, one might expect this to require $t_{\text{min \ transit}} \ge d$ for large $d$.  This is the case in $D\ge 5$ spacetime dimensions. There one again has $x_{*0} \approx d/2$, so it is clear that the higher dimensional analogue of our calculation will again give $\frac{t_{\text{min\ transit}}}{d} \rightarrow 1$. So in this sense, at least for $D\ge 5$, perturbative back-reaction on black hole spacetimes far from extremality comes close to saturating the theoretical bound on the shortest possible transit times for traversable wormholes\footnote{In a fixed four-dimensional asymptotically flat spacetime (not necessarily satisfying any positive energy condition) of total mass $M>0$, there is an infrared logarithmic divergence in the Shapiro time-delay for signals sent between distant points.  As a result, the fastest causal curve between such points always lies far from the center of mass, no matter what shortcuts might be available closer to this center.  This means that, even at large separation $d$, it is difficult to use arguments about causal curves connecting distant points to rigorously bound wormhole transit times.  It would be interesting to understand what bounds might be derived directly from the quantum focusing conjecture
\cite{Bousso:2015mna}.}. 

In contrast, the eternally-traversable MMP wormholes \cite{Maldacena:2018gjk} (related to our extremal limits) have $\frac{t_{\text{transit}}}{d} > 2$. Thus, while our non-extremal wormholes are more fragile and while they are traversable only for a limited period of time, for properly-timed signals they can be traversed significantly more quickly than corresponding MMP wormholes. This raises the interesting question of whether excited states of MMP wormholes might also have comparably shortened transit times for properly timed signals.  We leave such issues for future investigation.

\section*{Acknowledgements}
DM and ZF were supported in part by NSF grant PHY1801805 and in part by funds from the University of California. BG-W was supported by an NSF Graduate Research Fellowship.

\appendix

\section{Counterpoint:  Negative Energy causes collapse of cosmological wormholes}
\label{sec:dS}

In order to contrast with our analysis above, and also because it provides a convenient exactly solvable model, we now briefly discuss analogous computations for what one may call a cosmological wormhole.  Here we again consider a $\mathbb{Z}_2$ quotient of a covering spacetime $\tilde M$ having a globally-defined Killing symmetry, which in this case we take to be exact de Sitter space ${\rm dS}_d$.  In particular, in global coordinates we take the $\mathbb{Z}_2$ identification to be the antipodal map on the spheres at each global time. This is clearly a cosmological analogue of the $\mathbb{RP}_3$ geon.  It is thus natural to think of it as a cosmological wormhole, though we will not attempt to introduce a general definition of this term.

\begin{figure}[t]
\begin{center}
\begin{subfigure}{.49\linewidth}
\centering
\vspace{1.7cm}
\includegraphics[width = .6\linewidth]{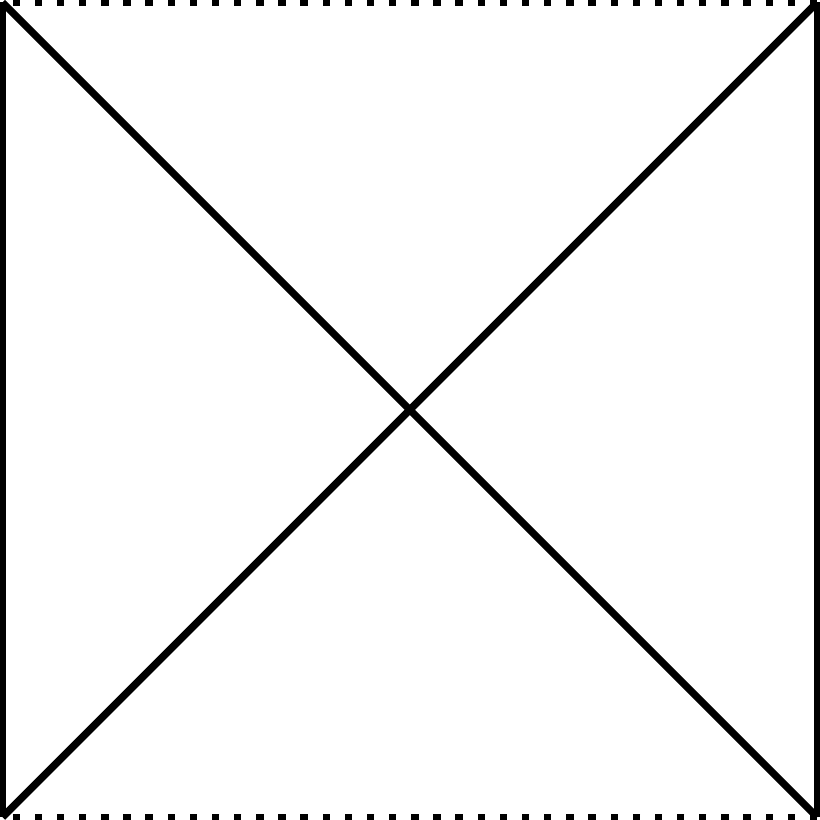}
\end{subfigure}
\begin{subfigure}{.49\linewidth}
\centering
\includegraphics[width = .6\linewidth]{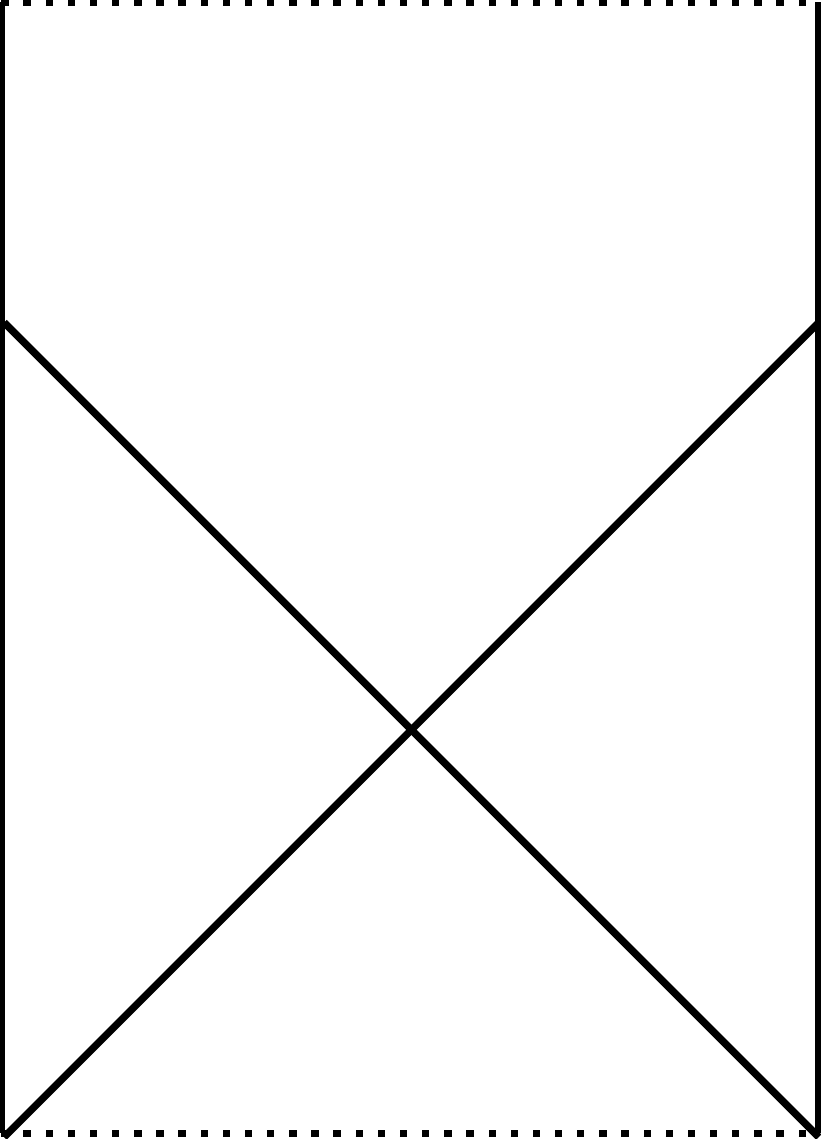}
\end{subfigure}
\end{center}
	\caption{(Left) A conformal diagram of ${\rm dS}_d$ showing time and the polar angle on the $S^{d-1}$ (the other $d-2$ angles are suppressed). The left edge is the south pole of the $S^{d-1}$ and the right edge is the north pole.  The diagonal lines denote light rays.  (Right) Perturbations satisfying the NEC generically make the diagram taller so that light rays can travel from the north pole to the south pole in finite time.  Here for simplicity we consider perturbations that preserve spherical symmetry. }
        \label{fig:dSfig}
\end{figure}

As is well known, in de Sitter space perturbations satisfying the NEC tend to make the conformal diagram taller so that -- at least in the natural sense defined by global coordinates -- wormholes become {\it more} traversable; see figure \ref{fig:dSfig}.  This is evident from the classic Einstein static universe solution, in which the addition of positive energy dust to an otherwise-empty de Sitter space removes the cosmological expansion and leaves a static cylinder that can be circled by causal curves arbitrarily many times.  That a similar effect occurs from general perturbations satisfying the NEC also follows from \cite{Gao:2000ga}. One thus expects the analogue of \eqref{DeltaV} to have the opposite sign.  And since periodic scalars in the Hartle-Hawking state should again violate the null energy condition, they should {\it not} make our cosmological wormhole traversable.  Indeed, they should make it more {\it non}-traversable than before.  All of these expectations will be explicitly realized below.

In particular, it is straightforward to compute the analogue of \eqref{DeltaV}, showing the effect of back-reaction.  For simplicity, we treat only the rotationally symmetric case.  The $\text{dS}_d$ metric with a general spherical perturbation is
\begin{equation}
ds^2=\ell^2\frac{-4dUdV+{{\left( 1+UV \right)}^{2}}d\Omega^2_{d-2}}{{{\left( 1-UV \right)}^{2}}}+{{h}_{UU}}d{{U}^{2}}+2{{h}_{UV}}dUdV+{{h}_{VV}}d{{V}^{2}}+h_{\Omega \Omega}d\Omega^2_{d-2},
\end{equation}
where $d\Omega^2_{d-2}$ is the standard metric on the unit $S^{d-2}$ and where $h_{UU}$, $h_{UV}$, $h_{VV}$, $h_{\Omega \Omega }$ are functions of $U$, $V$. On the horizon $V=0$, the linearized Einstein equation (with cosmological constant) yields
\begin{equation}
8\pi G {{T}_{UU}}=-\frac{d-2}{2{{\ell }^{2}}}\left( 2{{h}_{UU}}+U{{\partial }_{U}}{{h}_{UU}}+{{\partial }^2_{U}}{{h}_{\Omega \Omega }} \right).
\end{equation}
The negative sign in the above expression shows that positive null-energy gives a time-advance, while negative null-energy gives a time-delay.  In particular, we find
\begin{equation}
\Delta V = \frac{1}{8\ell ^2}\int ^{\infty }_{-\infty }h_{UU}dU=-\frac{\pi G}{d-2}\int ^{\infty }_{-\infty }T_{UU}dU.
\end{equation}

Since scalar two-point functions on $\text{dS}_d$ are known in closed form for any mass $m \ge 0$ and dimension $d$, we may dispense with any cosmic strings and simply study scalars on $M = \text{dS}_d/\mathbb{Z}_2$ coupled to pure Einstein-Hilbert gravity with a cosmological constant.  For simplicity, we ignore quantum effects from linearized gravitons.  Neglecting such contributions is justified in the presence of a large number $N$ of bulk scalar fields.

The scalar two-point function in the $\text{dS}_d$ Hartle-Hawking state (also known as the Bunch-Davies vacuum or the Euclidean vacuum) takes the closed form expression \cite{PhysRevD.12.965,PhysRevD.13.3224}
\begin{equation}
\begin{aligned}
G\left( {x,x'} \right) = & \frac{1}{{{{\left( {4\pi } \right)}^{d/2}}{\ell ^{d - 2}}}}\frac{{\Gamma \left( {\frac{{d - 1}}{2} - i\mu } \right)\Gamma \left( {\frac{{d - 1}}{2} + i\mu } \right)}}{{\Gamma \left( {\frac{d}{2}} \right)}}\\
& \times {}_2{F_1}\left( {\frac{{d - 1}}{2} - i\mu ,\frac{{d - 1}}{2} + i\mu ; \frac{d}{2}; 1 - \frac{{D\left( {x,x'} \right)}}{{4{\ell ^2}}}} \right),
\end{aligned}
\end{equation}
where $\mu \equiv \sqrt{m^2\ell ^2-\frac{1}{4}}$, and $D\left(x,x'\right)$ is the (squared) distance between $x$ and $x'$ in the $(d+1)$-dimensional Minkowski spacetime into which $\text{dS}_d$ is naturally embedded. In some references, $D\left(x,x'\right)$ is called the `chordal distance' between $x$ and $x'$.

In global coordinates, the de Sitter line element is
\begin{equation}
ds^2_d=\frac{\ell ^2}{\cos ^2\eta }\left(-d\eta ^2+d\theta ^2 + \sin^2 \theta d\Omega ^2_{d-2}\right),
\end{equation}
with $\theta \in [0,\pi]$.
The $\mathbb{Z}_2$ quotient identifies each point $\left(\eta ,\theta, \Omega \right)$ with point $\left(\eta , \pi - \theta , a(\Omega)\right)$ where $\Omega \in S^{d-2}$ and $a(\Omega)$ denotes the $S^{d-2}$ antipodal map.  We want to compute the two point function $G\left(U,U'\right)$ when the first point lies on horizon $\eta =\theta $ and has affine parameter $U$ and the second point lies on the image horizon  $\eta = \pi - \theta$ with affine parameter $U'$. The affinely-parametrized horizon is
\begin{equation}
\eta \left(U\right)=\arctan U,\text{ }\theta \left(U\right)=\arctan U,
\end{equation}
which becomes
\begin{equation}
T\left( U \right) = \ell U,{\rm{ }}X\left( U \right) = \ell ,{\rm{ }}\vec Y\left( U \right) = \ell U\vec z,
\end{equation}
in terms of the $d+1$ standard embedding coordinates $T,X,\vec{Y}$ such that $T^2 - X^2 -  |\vec{Y}|^2  = -\ell^2$. Here, $\vec{z}$ is a unit vector describing a $S^{d-2}$.
The affine-parametrized image horizon is
\begin{equation}
T'\left( U' \right) = \ell U',{\rm{ }}X'\left( U' \right) = -\ell ,{\rm{ }}\vec Y'\left( U' \right) = -\ell U'\vec z.
\end{equation}
Thus, the chordal distance is
\begin{equation}
D\left(U,U'\right)= - (T-T')^2 + (X-X')^2 + |\vec{Y}-\vec{Y}'|^2 = 4\ell ^2\left(1+UU'\right),
\end{equation}
and the two point function is
\begin{equation}
\begin{aligned}
G\left( {U,U'} \right) = & \frac{1}{{{{\left( {4\pi } \right)}^{d/2}}{\ell ^{d - 2}}}}\frac{{\Gamma \left( {\frac{{d - 1}}{2} - i\mu } \right)\Gamma \left( {\frac{{d - 1}}{2} + i\mu } \right)}}{{\Gamma \left( {\frac{d}{2}} \right)}}\\
& \times {}_2{F_1}\left( {\frac{{d - 1}}{2} - i\mu ,\frac{{d - 1}}{2} + i\mu ;\frac{d}{2}; -UU'} \right).
\end{aligned}
\end{equation}

The stress tensor is then
\begin{equation}
\begin{aligned}
{T_{UU}}\left( U \right) = & \mathop {\lim }\limits_{U' \to U} {\partial _U}{\partial _{U'}}G\left( {U,U'} \right)\\
= & \frac{1}{{{2^{d + 2}}{\ell ^{d - 2}}{\pi ^{d/2}}}}\frac{{\left( {d - 1 - 2i\mu } \right)\left( {d - 1 + 2i\mu } \right)}}{{d\left( {d + 2} \right)}}\frac{{\Gamma \left( {\frac{{d - 1 - 2i\mu }}{2}} \right)\Gamma \left( {\frac{{d - 1 + 2i\mu }}{2}} \right)}}{{\Gamma \left( {\frac{d}{2}} \right)}}\\
& \times \left[ { - 2\left( {d + 2} \right){}_2{F_1}\left( {\frac{{d + 1 - 2i\mu }}{2},\frac{{d + 1 + 2i\mu }}{2};\frac{{d + 2}}{2}; - {U^2}} \right)} \right.\\
& \left. { + {U^2}\left( {{d^2} + 2d + 1 + 4{\mu ^2}} \right){}_2{F_1}\left( {\frac{{d + 3 - 2i\mu }}{2},\frac{{d + 3 + 2i\mu }}{2};\frac{{d + 4}}{2}; - {U^2}} \right)} \right].
\end{aligned}
\end{equation}
Its integral is
\begin{equation}
\label{eq:gammas}
\int_{ - \infty }^\infty  {{T_{UU}}\left( U \right)dU}  =  - \frac{{\Gamma \left( {\frac{d}{2} + 1} \right)\Gamma \left( {\frac{d}{2} - i\mu } \right)\Gamma \left( {\frac{d}{2} + i\mu } \right)}}{{2{\ell ^{d - 2}}{\pi ^{d/2}}\Gamma \left( {d + 1} \right)}} < 0.
\end{equation}
Using the identity $\Gamma(1-z)\Gamma(z) = \frac{\pi}{\sin(\pi z)}$ one can rewrite the right-hand-side of \eqref{eq:gammas} as a $d$-dependent polynomial in $\mu$ divided by $\sinh(\mu + i \pi \frac{d}{2})$ (i.e., divided by either $\sinh(\mu)$ or $\cosh(\mu)$ depending on whether $d$ is even or odd).  The polynomial has a definite sign such that the overall expression is negative for all allowed $\mu$, and the factor of $\sinh(\mu + i \pi \frac{d}{2})$ in the denominator means that it decreases exponentially at large $\mu$.

\bibliography{bachweyl}
\bibliographystyle{utcaps}

\end{document}